# UC Santa Cruz
## Assessment

**Title**
Using Pre-/Post-Quizzes Intentionally in Curriculum Development and Evaluation

**Permalink**
https://escholarship.org/uc/item/6fz0181f

**Authors**
Cooksey, Kathy L
Jonsson, Patrik

**Publication Date**
2022-09-22

**Copyright Information**

# Using Pre-/Post-Quizzes Intentionally in Curriculum Development and Evaluation

Kathy L. Cooksey[*1] and Patrik Jonsson[2]

[1] Department of Physics & Astronomy, University of Hawaiʻi at Hilo, Hilo, HI, USA
[2] Hilo, HI, USA
[*] Corresponding author, kcooksey@hawaii.edu

## Abstract

Developing the final summative assessment of a course at the start of curriculum development is
an implementation of "backward design," whereby learning objectives are identified first and the
curriculum is engineered end-to-beginning to achieve them. We trained in backward design through
the Professional Development Program (PDP) and adapted PDP assessment ideas for evaluation of
curriculum designs and teaching efficacy. A pre-/post-quiz is an assessment administered the first
and last day of a course; a learner's scores are used to measure normalized gain: the ratio of what
a student learned during a course relative to what they knew entering it. The intentional process of
developing a pre-/post-quiz for every course focuses the educator on the essential understanding
desired of the learners exiting the course. The normalized-gain statistics for the course can then be
used to evaluate the course's efficacy, and improvements to the curriculum can be monitored by
tracking the normalized gains over time, using the same pre-/post-quiz. Moreover, an individual
instructor may self-evaluate their teaching efficacy by tracking normalized gains from all courses
over time. Here we discuss applying the practice of backward curriculum design starting with a
custom pre-/post-quiz and utilizing it for immediate and longitudinal evaluation, focusing primarily
on designing an entire undergraduate science course.

## 1. Introduction

"Backward design" is essentially being goal-oriented when developing curricula (Wiggins & McTighe, 2005). The use of "design" evokes an engineering process: developing, organizing, and assembling pieces to successfully accomplish the objectives. Identifying the course learning objectives (LOs) first enables a curriculum to be engineered to meet them; this is related to scaffolding but in a whole-course context. "Scaffolding" is the educational practice of pulling back on "the intellectual training wheels of cues, prompts, and tools" provided to learners "to see if [they] can perform with understanding of their own" (Wiggins & McTighe, 2005).

Backward design is an iterative process as the LOs are refined. Backward design was a core topic in the Professional Development Program (PDP), which was later run by the Institute for Scientist & Engi-neer Educators (ISEE) at the University of California, Santa Cruz (UCSC; see Hunter et al., 2010). Through the PDP, scientists, engineers, and/or educators learned about science education, inquiry teaching techniques (Metevier et al., 2022a, 2022b), and issues of diversity and equity in the sciences (Seagroves et al., 2022).

The basic PDP cycle was: (i) learn about science pedagogy at the workshop; (ii) begin designing an educational activity there with a team, for a specific venue; (iii) continue development after the workshop; (iv) implement the design at the venue; and (v) debrief about the design with specific discussion about what did and did not work. Many PDP participants would reenter the cycle by either improving the design and implementing it again or joining another design team; all the while, the PDP staff would elevate the training of returning participants.

With the built-in intention of improving designs and participants' teaching abilities, assessment was a fundamental theme of the PDP (Hunter et al., 2010; Metevier et al., 2022a, 2022b). Under ISEE, backward design evolved into "assessment-driven design", with a focus on developing the culminating assessment task and its rubric to assess learners (Hunter et al., 2022). In the PDP framework, *assessment* of learners can also serve in *evaluation* of the efficacy of the design and the educators themselves.

For example, one PDP venue, described in Cooksey et al. (2010), was a summer residential science, math, and enrichment high-school program at UCSC. PDP participants developed and used pre-/post-quizzes as formative and summative assessments. A conceptual pre-quiz given at the start of a learning module would formatively assess the starting level of the learners and inform the educational approach. The same conceptual post-quiz administered at the end of the module would assess what was learned by the students; this can be quantified as the normalized gain (Hake, 1998):

$$g = \frac{\text{Post [\%]} - \text{Pre [\%]}}{100\% - \text{Pre [\%]}},$$

where "Pre/Post [%]" are the percentage scores on the pre-/post-quizzes, respectively (Eqn. 1). The normalized gain is the ratio of what a student learned to what they knew entering; it supported the educators evaluating their design and teaching efficacy.

Using normalized gain to assess student learning is a standard practice in physics-education research (e.g., Hake, 1998). There are researched pre-/post-quizzes available, such as the Force Concept Inventory for introductory physics (Hestenes et al., 1992); the Light and Spectroscopy Concept Inventory for introductory astronomy (Bardar et al., 2006); and the Quantum Mechanics Conceptual Survey for modern physics (McKagan et al., 2010). "Researched" refers to how the questions were developed and validated, as well as to using normalized gains measured with the pre-/post-quizzes to evaluate teaching practices across classrooms.

It may work to use an already developed pre-/post-quiz in the process of backward designing a curriculum; it essentially means the course LOs are defined by others because, in backward design, the assessments are based on the LOs. These assessments could in principle be anything that elicits evidence of the learners' mastery of the LOs; here we describe multiple choice pre-/post-quizzes as the assessments.

In this article, we detail developing a bespoke pre-/post-quiz for a course as part of the backward-design process; the pre-/post-quiz is crafted in alignment with the course LOs. We describe implementing this practice in a university setting, primarily focusing on designing an entire undergraduate science course. First, a brief note about roles. Cooksey is the full-time educator responsible for curriculum development and instruction; Jonsson is consulted as necessary but final decisions and actual teaching lay exclusively with Cooksey. In this article, "we" is used collaboratively with the understanding of these roles.

This article is organized as follows: §2 details developing pre-/post-quizzes in a backward-design context; using the pre-/post-quizzes to assess learners and evaluate a course design and teaching efficacy is discussed in §3; and §4 is a summary.

# 2. Backward design via developing pre-/post-quiz

## 2.1. Identifying learning objectives

A full discussion on how to develop course LOs (e.g., Chapter 3 of Wiggins & McTighe, 2005) is beyond the scope of this article. We simply make four points about LOs before discussing pre-/post-quiz development in §2.2.

First, we consider two perspectives on the LOs of an educational design: the learners' and the educators'. Often, the learners are content-oriented and engage with the course or activity to learn about the topic (e.g., astronomy); the educators also likely desire learning of content. In addition, they may be process- and/or motivation-oriented. "Process" refers to learners experiencing (and ideally improving in) the practices of a field: asking investigable questions, problem solving, data analysis, etc. (Institute for Inquiry [IfI], 2006). Examples of motivation-oriented LOs are: engendering a sense of accomplishment in the learners so they are empowered to pursue the subject further or fostering an appreciation of the field, so learners support it as taxpaying citizens.

Second, a general rule of thumb for articulating LOs is: they should include the evidence or assessment for achievement. For example, "form a conceptual framework of the content, structure, and evolution of the universe" might be an introductory-astronomy LO; it does not indicate what is evidence of accomplishing this. An improved phrasing might add "as evidenced by the ability to connect topics in astronomy in multiple, meaningful ways," or perhaps the educator revises so this one LO becomes several: "as evidenced by the ability to: use correct terminology for astronomical objects; correctly describe which objects are larger/smaller and/or components of other objects; and correctly describe how astronomical objects change over time."

To assess whether learners master rich LOs like the previous example, the educator might develop an assessment task within the course where the learner produces a concept map or there is a free-response question about connecting two topics not explicitly linked in the course. The educator, in having dissected and revised the LOs, would be well positioned to develop a rubric that identifies the dimensions of the assessment task and articulates what mastery looks like. This process is an example of assessment-driven design as taught by the PDP under ISEE (Hunter et al., 2022).

Third, developing and implementing assessments of whether the LOs are achieved is integral to backward design; if a LO is not worth assessing, it is arguably not worth being a goal of the curriculum. The assessment may be a task or assignment for the learners. It may be more informal feedback such as discussions with learners or course evaluations; for example, a LO may be students connect the course content to their everyday life, and the evidence of success comes from them spontaneously sharing the connections they identify.

Likely, all content-oriented and some process-oriented LOs can be assessed with a well-crafted conceptual pre-/post-quiz. This is addressed further in §2.2.

Fourth, developing LOs is an iterative process in backward design; they are intentionally revised as the curriculum emerges. The interplay between identifying a LO and reverse engineering the course to achieve it can lead the educator to realize the starting point (e.g., the pre-requisites of the course), the logistical constraints of the course (e.g., length of term, required topics), and a pedagogically sound scaffolding sequence do not lead to achieving the desired LO, in which case, perhaps the LO is de-scoped or other LOs must be removed or revised.

## 2.2. Pre-/post-quiz principles

In §2.3, we describe a process for crafting a be-spoke conceptual, multiple-choice pre-/post-quiz while backward designing a curriculum. In this section, we address why conceptual and why multiple choice; we also provide basic instructions for the quizzes and guidelines for implementing them effectively in a course.

We define "conceptual" to include some types of quantitative (computational) questions, so long as the math and content knowledge required are clearly within the scope of the course pre-requisites and/or LOs; we will refer to non-conceptual questions as "applied." For example, we consider the following to be a conceptual question:

> The flux of an object is an inverse-square law. Thus, if the flux of the Sun at Earth is 1361 W m$^{-2}$, what is the flux at Mars, which is nearly twice as far from the Sun as Earth?

The answer is 1361 W m$^{-2}$ divided by two-squared, and the choices are numerically well spaced, so minor calculation errors do not affect the assessing power. In fact, like the alternative, incorrect choices on the Mechanics Baseline (Hestenes & Wells, 1992), the other choices for the above question are selected to probe common student misapplications of an inverse-square law. Thus, the boundary between a quantitative conceptual question and an applied one is subjective; in the spirit of backward design, the discriminators are the LOs: if a quantitative question assesses minutia, trivia, or one-offs from the course, it is a poor choice for a pre-/post-quiz.

There is evidence that sound conceptual under-standing is a necessary though insufficient condition for applied questions (Hestenes & Wells, 1992). The latter would require either providing equations, if the assessment is of "can the learner use them properly" or expecting the learner to know the equations *and* use them properly, so questions may be assessing two possibly different objectives (content and process, respectively).

An ideal pre-quiz has no learner getting a zero or a perfect score, as either extreme is loss of information — if zero, what does the learner know?; if perfect, what does the learner not know? A course where most students score very low or very high on the pre-quiz would need to be rapidly modified to intentionally account for the formative-assessment result. For reference, a multiple-choice pre-/post-quiz has a random-guessing score that is the sum of the reciprocals of the number of choices of all questions. For example, if there are 30 questions with five choices each, the random-guessing score is $30 \times 1/5 = 6$ or 20%. Scores significantly below the random-guessing score may indicate substantial misconceptions, especially if the incorrect choices are well chosen.

Multiple-choice pre-/post-quizzes are common in physics-education research (e.g., Hestenes & Wells, 1992; Bardar et al., 2006), but here we focus on developing bespoke multiple-choice pre-/post-quiz-zes in this article for two main reasons. First, a multiple-choice pre-/post-quiz is faster to grade, analyze, and, thus, react to; the latter of which is the purpose of formative assessment. Second, each question can be keyed to LO(s) and/or concept(s); even the alternative, incorrect answers can be keyed by "commonsense" misconception (e.g., Hestenes et al., 1992). In evaluating a course design post-in-struction, the ability to sort questions by LO-key and analyze in detail is important when considering improvements to the LOs and curriculum (see §3.1).

Another advantage of a custom pre-/post-quiz over one from the literature became apparent in the COVID-19 pandemic: the instructor controls the in-person or online format and delivery of the quizzes, whereas authors of other assessments tightly control them for security and validity. Thus, at UH Hilo, the pandemic disrupted the Department's lon-gitudinal use of the Mechanics Baseline (Hestenes

& Wells, 1992), but did not disrupt the use of bespoke pre-/post-quizzes.

Free-response questions can be used for pre-/post-quizzes, and they have the significant benefit of providing insight into learners' thinking. They require well-developed rubrics to assess answers accurately with respect to LO(s) and consistently across learners; rubric design is a rich, important educational tool (e.g., Chapter 8 of Wiggins & McTighe, 2005), a significant part of ISEE's assessment-driven design (Hunter et al., 2022), and beyond the scope of this article. Free responses are also useful for developing robust multiple-choice questions since common misconceptions can be used in the alternative, incorrect choices; this is a useful approach when developing a pre-/post-quiz for education research.

It helps if learners simply *taking* the pre-/post-quizzes were part of the course grade; in our experience, 2.5% per quiz provides sufficient motivation. The educator is invested in having the pre-/post-quiz results (as unbiased as possible) and modest credit for effort "pays" for that. Having the quizzes for credit only also minimizes the stress for the students, which increases their chance of peak performance.

Ideally, the pre-quiz is given at the top of the first period[1] and the post-quiz, on the final day of instruction, which is not necessarily the end of the learning since students often will then study for the final exam. Greeting new learners with a quiz might give a negative first impression. It helps alleviate discontent to emphasize verbally how the pre-quiz is for credit only, just for a good-faith effort, and how it helps set the starting level of the course to maximize effectiveness. After everyone is done, it also helps to explain how the content of the pre-quiz represents essential LOs of the course, so the students now have an overview; this can be a segue to reviewing the syllabus where the LOs are listed.

Finally, we emphasize here that any assessment only probes what a learner demonstrates, which is a proxy for their true understanding. For example: did the learner answer incorrectly because they misunderstood the concept or misread the question?; if free-response, is the muddled explanation due to lack of understanding or unclear writing?; or does the incorrect numerical result indicate a minor calculation error or a major misunderstanding?

## 2.3. Developing custom pre-/post-quizzes

As a first step in the backward design of a curriculum, the educator drafts LOs (see §2.1). If they do not know the course content to the level at which they can do this, they might try compiling a repository of conceptual questions. We find the process of building a repository helps review the course content, organize it topically, and begin prioritizing it, which supports drafting LOs. The questions are used in the bespoke pre-/post-quiz and throughout the course for formative assessments (e.g., think-pair-share; Mazur, 1997) and summative assessments (e.g., assignments, tests).

Once draft LOs exist, the pre-/post-quiz can be drafted, perhaps culled from the question repository. Again, we emphasize: likely, all content-oriented and some process-oriented LOs can be assessed with a well-crafted conceptual pre-/post-quiz, and it is to these LOs we refer in the context of the pre-/post-quiz. Other LOs — and even some which the pre-/post-quiz assesses — may be assessed via more detailed assessment tasks and their associated rubrics, as mentioned in §2.1 and proscribed by ISEE's assessment-driven design (Hunter et al., 2022).

A question assessing a process-oriented LO may ask the learner to extrapolate from what they were explicitly taught to a connected context; for exam-ple, the flux question in §2.2 can be used for an introductory-mechanics pre-/post-quiz if Newton's law of universal gravitation (also an inverse-square law) is emphasized.

---

[1] The bespoke pre-/post-quizzes can be given as take-home or online assignments, so long as the instructor is comfortable with students possibly not taking them as seriously or cheating and with the security concerns of the questions being saved and distributed.

While discerning LOs and associated pre-/post-quiz questions, we find it fruitful to sort related equations between an official equation sheet and a "content not on the equation sheet" section of the syllabus.[2] Crystallized intelligence (i.e., memory) is necessary for learning and understanding (Medina, 2014); such knowledge is foundational "content not on the equation sheet." Fluid intelligence is the ability to adapt, reason, and problem solve; the equation sheet provides specific tools to adapt crystallized knowledge in new contexts. Essentially, explicitly determining what is *not* on the equation sheet is an exercise in articulating what content LOs are so foundational that learners must crystallize them in memory. With a clear division between foundational knowledge and other content, quantitative conceptual questions can be discerned from applied questions (as defined in §2.2).

We find it crucial to write detailed solutions to the emerging pre-/post-quiz. Explaining the answers as if for a learner forces the educator to reflect on the desired evidence of understanding, and this cycles back to aligning the assessment with the emerging LOs and/or revising the LOs to focus on the actual pedagogical intent of the course. In addition, if the pre-/post-quiz were not multiple choice, the detailed solutions would be a first step to developing a rubric to assess student responses with respect to LO(s) and consistently across the learners.

The time allotted for the pre-/post-quiz should be sufficient for everyone to reasonably complete it, thus the number of questions scales with how much class time the educator is willing to devote to it. There needs to be sufficient questions to span the LOs to be assessed; likely, LOs should correspond to more than one question, especially if there is intention to analyze the LO-keyed questions. Sufficient questions also support the broad-stroke assessment of measuring normalized gain to be statistically significant. We find 25–30 questions for 40–50 minutes works well.

The phrasing of questions should be carefully considered; this is as simple as double-checking so e.g., a missing "not" does not upend the question. It is as complex as ensuring the question provides sufficient information to be assessing what the educator wants it to. For example, the flux question in §2.2 probes a different LO if the prelude about flux being an inverse-square law is not provided. With the first sentence, the question focuses on learners implementing the scaling of an inverse-square law; without it, learners must know flux is an inverse-square law *and* apply it.

Phrasing is also important in the context of equity and inclusion. For example, McCullough & Meltzer (2001) reworded the Force Concept Inventory (Hestenes et al., 1992) to use more "daily-life" or "female-oriented" contexts and demonstrated it elevated female scores on some questions. A very specific illustration is avoiding mechanics questions about frictionless motion on ice at UH Hilo, where a significant fraction of students may have zero experience with ice.

Once the synergistic development of the LOs and pre-/post-quiz is nearly finalized, it is extremely valuable to find at least one person (e.g., colleague, advanced student) to test-drive it and provide feedback if willing; what they answered correctly versus incorrectly and how long they took are the basic information. Jonsson has sound physics and astronomy training and intuition, as well as an understanding of science pedagogy and PDP practices, and regularly serves as Cooksey's guinea pig. His feedback contributes to the revision of the pre-/post-quiz, not just for general clarity but also to better focus on what is intended to be assessed; sometimes questions are revised or replaced with more targeted ones. Students who have taken the course are good beta testers; discussing with them about their incorrect answers can also inform revision and/or replacement of questions.

---

[2] For example, see PHYS170 and PHYS272 syllabi and equation sheets at http://guavanator.uhh.hawaii.edu/~kcooksey/teaching/UHH/UHH.html.

All feedback on the pre-/post-quiz may lead to modifications to the LOs and course design as well. For example, the educator may realize that a LO assessed by some questions an advanced student answered incorrectly may be even more important to cover carefully in the course or that the LO is minutia that does not really matter down the line.

Test-driving the pre/post-quiz well, hopefully, reduces or eliminates the need to discard questions upon realizing there were typos or ambiguities. Hestenes et al. (1992) rejected a couple questions from the Force Concept Inventory after researching its use in various high-school and undergraduate classrooms because of phrasing; they realized poor reading comprehension led to misunderstanding.

The pre-/post-quiz may need to be revised after the first application if some or all LOs are sufficiently redesigned. This affects longitudinal evaluation of curriculum and instructor efficacy as discussed in §3.2. At a minimum, revising or replacing questions increases the noise in such evaluation.

However, educators may be committed to the LOs as stated, in which case, the educator may interpret poor performance on a consistent portion of the pre-/post-quiz as an indication to improve their teaching and/or curriculum to better support the intended LOs. Improvement can be monitored with a longitudinal assessment (see §3.2). For example, we included Newton's universal law of gravitation in the introductory-mechanics LOs and custom pre-/post-quiz. We then tracked this LO over four iterations of the course, tightening the course design, noting students' particular struggles, and trying to better support this particular LO.

"Teaching to the test," of course, makes the related summative assessment suspect: was learning achieved or just rote memorization? First and foremost, we emphasize that the purpose of backward design is to align the assessments with the LOs, thus the teaching and learning are aimed toward the LOs. This is not "teaching to the test" but "testing for what is taught."

Second, one benefit of a conceptual-question repository (described earlier) is a wide selection of questions from which to draw upon for formative and summative assessments within the curriculum, and this can help mitigate the negative accusations of "teaching to the test." Some questions will be excellent analogs to ones used on the custom pre-/post-quiz. Even if questions from the pre-/post-quiz are used verbatim, learners may not discern which are pre/post ones if the course is rife with conceptual questions. Also, multiple-choice questions can easily be morphed into free-response ones, which supports learning and also masks the pre/post questions. More importantly, learners get ample practice with multiple-choice conceptual questions (a process skill) and, as stated previously, our starting axiom is that conceptual understanding is paramount.

# 3. Using pre-/post-quiz results for assessment and evaluation

The theme of this article is using pre-/post-quizzes intentionally, from designing a curriculum backward, evaluating success, and redesigning for improvement. The pre-/post-quiz can simultaneously be a tool to *assess* learners and to *evaluate* course and teaching efficacy. Evaluation of a course design is mixed up with evaluation of its implementation (i.e., one instance) that itself is tied to instruction (i.e., an individual's teaching practice). As we discuss immediate and long-term evaluation, we endeavor to be clear to what the evaluation refers.

We focus on evaluation an instructor can do on their own, with minimal resources. We are not describing education research,[3] for either validating a pre-/post-quiz (e.g., Hestenes et al., 1992, Bardar et al., 2006) or proving one course design leads to better outcomes than another (e.g., Hake, 1998, Dori & Belcher, 2005). We assume most instructors do not have the resources (which includes time!) to rigorously validate teaching tools or teach a true control class. However, a curriculum's normalized gains, as measured by the bespoke pre-/post-quiz, can improve with deliberative improvement in design and/or implementation (as described in §2.3 and below), and an educator can evaluate how changes (major and minor) in curriculum, implementation, and/or instruction over time affect outcomes.

The backward-design process detailed in this article *evolved* during our teaching experience, beginning with the PDP and through teaching at UCSC, MIT and, primarily, UH Hilo. Some practices described here, like keying pre-/post-quiz questions by LO for detailed assessment and evaluation, we only recently learned to do. To date, we have not keyed pre-/post-quizzes from all previous courses to conduct the analyses suggested in this section.

## 3.1. Immediate assessment and evaluation

As mentioned previously, a pre-quiz given at the start of a learning module formatively assesses the starting level of the learners and informs the educational approach. If every question were keyed to at least one LO and/or topic, the instructor would have a rough guide to what the class understands as each assessed LO/topic is introduced in the course and can adjust accordingly; an instructor with sufficient experience teaching the material is more likely to successfully scaffold the students' learning from the assessed to desired level. Any instructor, but particularly a more novice one, would benefit if the alter-native, incorrect answers were also keyed by "commonsense" misconceptions; such higher-resolution assessment would improve scaffolding.

During a course, as each pre-/post-quiz-related LO is covered, the instructor can use the questions verbatim or their repository analogs (see §2.3) for intermediate formative and summative assessments. In one of our basic "learning modules," learners: (i) are introduced to a concept via reading, lecture, etc.; (ii) practice with the concept via an assignment, experiment, etc.; (iii) receive feedback (e.g., grading, posted solutions); and (iv) are given a summative assessment. Learners are formatively assessed in steps (i)–(ii), and any reactions to improve learning outcomes occur via step (iii) and e.g., lecture before step (iv).

Practicing backward design, each learning module's summative assessment is designed before developing steps (i)–(iii). So LOs are assessed by the pre-/post-quiz, the final exam, intermediate tests, assignments, and class time (e.g., think-pair-share, problem solving) — all have questions tightly focused on the LOs. Analogous questions from the repository and free-response versions of the pre-/post-quiz questions are used throughout a course. In practice, learning modules are interwoven, though students might complain about learning new topics before a test that does not cover them.

After a summative assessment, we present a table that links each test question to content covered via assignment and lecture (indicating where the content was in the assigned reading is unnecessary, since reading assignments tend to cover more than the true focus of the course.) Such a table is also a sound *evaluation* of the curriculum being designed backward.

---

[3] Educators interested in education research need to review their institutions' policies and procedures for human-subjects research. For reference, the data in Figures 1–2 were collected as an intentional practice of our teaching approach. To ethically present them here, we secured a waiver for the data in hand and committed to procuring consent from future students.

Using the pre-quiz for the post-quiz at the end of a course enables assessing learning via the normalized gain (Eqn. 1). The class's median normalized gain is a broad-stroke assessment of course efficacy. Hake (1998) determined an average gain of 0.25 was the dividing line between effective and ineffective instruction. The study compared traditional, lecture-based courses with ones using significant active-learning strategies. Essentially, the best traditional courses maxed out with gains of 0.25 while even fairly novice implementations of active-learning strategies achieved 0.25 and the best, up to 100% (i.e., students learned everything *assessed*).

Similarly, we adopt a *median* class gain of 0.25 as a holistic indication of a successful course (design and implementation); a median is less susceptible to outliers than an average and more robust for smaller classes. Note: negative normalized gains suggest guessing but could indicate a student uninfluenced by a course or confusion introduced by the course if the negative gain is statistically significant. For reference, from Monte-Carlo simulations of classes where all students guess, the maximum median normalized gain is roughly 0.1.

We share the normalized gain results with the students. When final grades are posted, individual normalized gains are disclosed privately to the learners, as well as the class's median normalized gain to everyone. The results are explained as follows:

> In physics-education research, there is a quantity called "Normalized Gain" which equals (Post [%] – Pre [%])/(100% – Pre [%]), where Pre and Post are the grades on the same conceptual quiz, given before and after the course. The normalized gain measures how much a student learns (Post – Pre), relative to how much they could have learned, given what they knew starting off (100% – Pre). The best traditional lecture-based courses have gains around 25%. Active-learning strategies (such as our in-class concept questions and problem solving) have gains of at least 25%, and the best have gains of 100% (i.e., 100% understanding).

If applicable, we additionally specify, "Negative gains indicate guessing," typically just to the learners with the negative results. Students who did not complete the pair of quizzes receive estimated normalized gains, adopting the class median on the missing quiz.

As with the pre-quiz formative assessment, analysis can be more detailed and pinpoint design elements that were successful and ones that need revision. Questions with a statistically significant increase in learners being correct on the post-quiz than pre-quiz indicate success with the LOs associated with those questions. (It does not have to be every learner who was correct on the pre-quiz remains correct on the post-quiz; random guessing is a factor.)

If there were questions with the opposite behavior, the instructor should consider whether they were flawed (e.g., phrasing) and otherwise reflect on what might have led to the consistent shift to misunderstanding. In this situation, if the incorrect answers are keyed by "commonsense" misconceptions, details of what learners shifted from and to may illuminate what happened. In a similar vein, examination of questions without statistically significant change in correctness frequency is useful if the incorrect answers are keyed; perhaps, at least, understanding in what direction the needle moved (if consistent) can inform redesign and future implementation.

Even after the post-quiz, alternative phrasings or repository analogs can be used in the final exam and provide a second normalized-gain measurement (and any subsequent analysis available if questions and alternative choices are keyed). There is more uncertainty with this assessment because an alternative phrasing or analog question may not probe the same LO/content. On the other hand, the small shift in perspective on a pre-/post-quiz question sheds light on the robustness of a learner's understanding; a learner answering the original post-quiz and final-exam alternative questions correctly is highly suggestive of robust understanding. Other benefits of a post-quiz embedded within the final exam include: (a) an indication if students used the post-quiz experience to guide their studying for the final exam and (b) possibly some ability to address the "teaching to the test" argument. For these reasons, and because we prioritize conceptual understanding, our final exams are 30–60% conceptual (the rest, applied).

We emphasize: the objective of the evaluations enabled by a pre-/post-quiz is to reflect upon and intentionally improve a course design, with implementation and/or instruction being affected as needed to support learners and learning.

## 3.2. Longitudinal evaluation

We advocate development and implementation of pre-/post-quizzes as part of the backward-design process for all curriculum (re)development; we also know most instructors teach different courses over time. Therefore, here we discuss long-term evaluation of course design, implementation, and instruction, using the heterogeneous data a backward-design-practicing educator may naturally collect.

First, an instructor may monitor improvement in one course by tracking median normalized gain, at the top level, and more detailed metrics (e.g., LO-by-LO, aspirational LOs from §2.3). Why any evaluation metric changes from one course to the next will likely be imprecisely known because: the learners are different; the instructor is more experienced; there are minor or major revisions to the course design; the time the course is offered is different; etc. The goal, however, is to detect improvement — generally increasing normalized gains and/or more LOs being consistently mastered — over several iterations. Examples of median normalized gains from our courses are presented in Figure 1.

Second, expanding on this, an instructor may self-evaluate their teaching efficacy by monitoring median normalized gains across all courses over time.

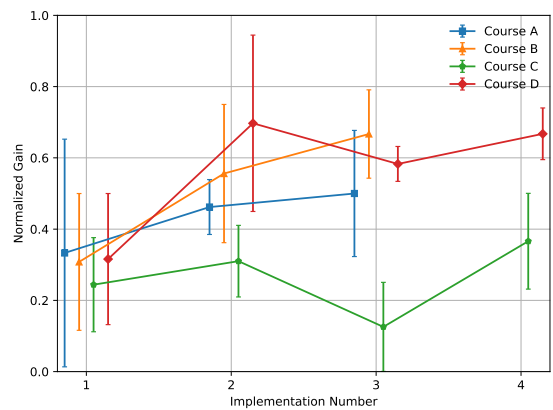

**Figure 1: Normalized gain over time for repeatedly implemented course designs by the same instructor.** Data are from four courses for which we developed the curriculum and pre-/post-quizzes and were implemented (i.e., taught) three or more times, between spring 2014 and fall 2021; class sizes ranged from six to 54. The symbols are the median normalized gain (Eqn. 1) for the class, offset horizontally for clarity; the error bars show the median absolute deviation. Generally, the median gains are greater than 0.25, which indicate effective designs and implementations; also, Course B shows general improvement between the first and third implementation, considering the error bars.

This is a heterogeneous dataset; however, the non-uniformity can be reduced if each course's pre-/post-quiz assesses learners over the same relative dynamic range. With that caveat, adopting 0.25 as the minimum gain of an effective course (Hake, 1998), a solid instruction record would be consistently hitting or exceeding normalized gains of 0.25, with a positive trend of gains increasing with time, especially when repeating the same curriculum. If a curriculum were becoming more ambitious, a steady (flat) normalized gain would indicate a solid instructional record.

Third, an instructor may assess retention of learning through a course sequence or even beyond. For example, we used some of the same pre-/post-quiz questions in a two-semester introductory astronomy courses for majors and in a higher-level course for which the introductory courses are pre-requisites;

all three courses had overlapping LOs, so reusing the same questions was pedagogically sound. With the same pre-/post-quiz questions, one can examine if the students tended to grasp the associated LOs in that first semester and retain them or whether the understanding oscillated for part or all of the sequence. Once again, interpretation of *why* is difficult because of the numerous variables (e.g., duration between courses). Of course, the tracking of learning retention is best served if all instructors within a program (e.g., major) agree to and use the same subset of pre-/post-quiz questions, ideally these would represent programmatic LOs.

Fourth, with more information, an instructor may evaluate fairness. For example, Figure 2 shows the median normalized gain as a function of letter grade for 309 students, from 21 courses taught over 12 semesters. Briefly, final course letter grades were assigned by sorting the students by their cumulative percentage of points and assigning letter grades in ranked order, attempting to balance demonstrated mastering (i.e., cumulative percentage) and improvement (via general positive performance trend and good normalized gain). To mitigate personal biases, a learner with a lower cumulative percentage never received a higher letter grade than a learner with a higher percentage.

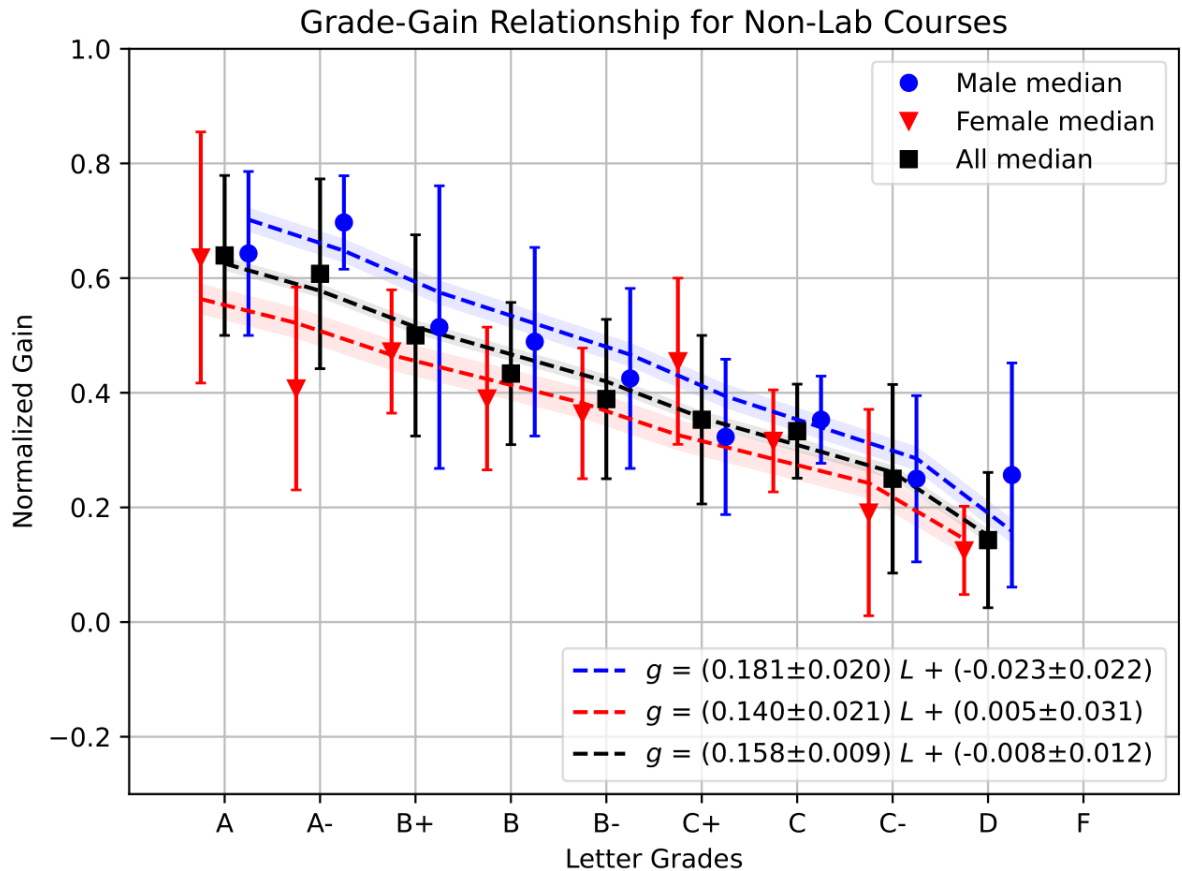

**Figure 2: Normalized gain as a function of final course letter grade for 309 learners** (138 female, 171 male) from 21 of our non-lab physics and astronomy courses over 12 semesters (fall 2014 to fall 2020, excluding spring 2018). These learners completed the course, took the post-quiz, and did not receive an F. The sex assignations are Cooksey's and not necessarily what the students would identify as (see §3.2 for why). The median gain *g* for all learners of each letter grade is shown as a black square; the median results for female and male learners are the filled red triangles and blue circles, respectively, with small horizontal offsets for clarity. The error bars reflect the median absolute deviation. Linear fits to *g* as a function of *L*, GPA equivalent of the letter grade (e.g., A = 4.0, A– = 3.7, B+ = 3.3, B = 3.0, etc), are shown with dashed lines and shading for the ±1-σ uncertainty; the fit results are given in the legend. Inferences from this figure are detailed in §3.2.

What is primarily learned from an analysis like Figure 2 is that even students with low final letter grades can (and do) gain (i.e., learn). Learners with higher final grades tended to learn more. It is difficult to disentangle causation from correlation since letter grades were assigned with knowledge and attention to normalized gains, so that learners with relatively high gains (compared to their classmates with close cumulative percentages) tended to be bumped up to a higher letter grade.

A secondary but very limited analysis is any possible bias with respect to learner sex. We were motivated to consider this upon learning how the same grade can affect women differently than men (e.g., a B may discourage a woman while a man may be neutral or encouraged; Marshman et al., 2018). However, by this time, years of data had already been collected and there was no well-justified or consistent way to ask previous students for their gender. The terms "male" and "female" here and in Figure 2, though not ideal, are used as shorthand to refer to the gender we perceived. We acknowledge that sex and gender are different concepts, that neither is necessarily binary, and that our perception of a person's gender presentation is a poor substitute for their own declaration.

For the entire sample in Figure 2, the median gain is 0.42±0.18 overall, 0.38±0.19 for female learners, and 0.45±0.18 for male, where the uncertainties are the median absolute deviations. There is a difference, but it is not statistically significant.

Similarly, there is a slight difference in the gain as a function of letter grade in Figure 2; the slope is shallower and intercept is higher for female learners. Again, the difference is not statistically signifi-cant. However, it could be the female students typically earned their final letter grade by consistent effort during the course (i.e., accumulating percentage points and demonstrating consistent or improving grades), as opposed to a test-based metric like normalized gain.

It is also possible there is something in our course design and/or teaching that is less effective for female learners. To possibly close the gap, there are a few approaches we could take in redesigning curricula. On the low-effort side, we could more formally use values affirmation (Miyake, 2010); this would require asking the learners to express their values more than once (as is the current practice), early in a course. An intermediate redesign effort would be developing and incorporating projects where learners connect the course content to their majors (Benderly, 2013) and/or a community problem (Margolis et al., 2000). A more targeted approach would be to qualitatively explore male and female experiences and implement changes based on the feedback.

There are more evaluations an instructor may conduct with the pre-/post-quiz grades and data collected naturally in a course. For instructors who need to apply for contract renewal, tenure, and/or promotion, the evaluations and analyses detailed in this section may provide solid evidence of teaching efficacy as well as sound pedagogical practices.

When presenting analyses such as Figure 2 in a tenure/promotion dossier, possible "teaching to the test" accusations can be combatted by pointing out how rare it is a learner gets 100% on the post-quiz (only three of the 309 in Figure 2). We find that backward design, intentional reflection and redesign, and aspirational LOs require explicit explanation in the dossier. Another approach would be to demonstrate the normalized gains from a researched pre-/post-quiz correlate to those from a custom one; this requires tracking both results for several iterations of a course. Such data may also support an educator's education research if their aim is to also produce a validated pre-/post-quiz.

# 4. Conclusion

We described a framework for developing curricula with sound pedagogical underpinnings. The essential steps are:

1. With the content landscape in mind, draft course learning objectives (LOs; §2.1); reviewing the course content by compiling and organizing a concept-question repository (§2.3) may be helpful

2. Use LO-assessing questions to develop a conceptual, multiple-choice pre-/post-quiz (§§2.2–2.3)

3. Once LOs are relatively fixed, design the curriculum backwards to achieve them

4. Implement the course, using the pre-quiz for formative assessment, intermediate conceptual questions for formative and summative assessments, and the post-quiz for summative assessment (§3.1)

5. Reflect on the curriculum design as part of evaluation (and experience) and revise as necessary (§3)

Within these barebones steps are revise, revise, revise — with attention to the goals. It is important to consider some objective evidence when reflecting on the efficacy of a curriculum design, and we described how using a pre-/post-quiz to assess student learning readily provides evidence for evaluation purposes. The more sophisticated uses of a pre-/post-quiz to elicit evidence and evaluate efficacy, as described in §3, is of secondary importance. We emphasize: just because one cannot do everything does not mean one should not do something.

Notably lacking in this article are details on step 3: actually designing the course. The main message is that each element of a curriculum should have a rationale behind it, like a move in chess. Accommodating logistical constraints can be a reason behind design choices, so long as support for learning is maximized within the constraints: class size; length of class periods; length of term; (no) funds for supplies or equipment; classroom layout; etc.

Otherwise, curriculum designs are heavily influenced by personal style; for this reason, it is often difficult to adopt a course design and materials without revising them. Some recommendations for curriculum development are interspersed in this article, such as intentionally documenting the division between content on an equation sheet versus not in §2.3 and our typical learning modules in §3.1. Both examples are grounded in the principles of backward design.

Backward design is a powerful tool for successful (i.e., pedagogically sound) curriculum development. It also forces the educator to articulate their intentions with each element of the course design. Sharing as much of the pedagogical rationale behind curriculum elements with the learners is a good way to foster their trust and increase the chances they engage successfully with the learning process.

# Acknowledgements

We acknowledge Lisa Hunter and ISEE generally for facilitating us learning about science pedagogy and issues of diversity and equity in the sciences. We are grateful for the guidance and input of our editor Scott Seagroves.

The PDP was a national program led by the UC Santa Cruz Institute for Scientist & Engineer Educators. The PDP was originally developed by the Center for Adaptive Optics with funding from the National Science Foundation (NSF) (PI: J. Nelson: AST#9876783), and was further developed with funding from the NSF (PI: L. Hunter: AST#0836053, DUE#0816754, DUE#1226140, AST#1347767, AST#1643390, AST#1743117) and University of California, Santa Cruz through funding to ISEE.

KLC appreciates the support of the University of Hawai'i Institutional Research Board and thanks all her students over the years that provided the data used here, not to mention the experience and feedback that has helped her improve.